\documentclass[a4paper,11pt]{article}
\usepackage{graphicx,amssymb,amsmath,amsthm,textcomp}
\input{psfig.sty}
\usepackage{color}
\newtheorem{theorem}{Theorem}
\newtheorem{lemma}{Lemma}

\newtheorem{definition}{Definition}
\newtheorem{observation}{Observation}

\newcommand{\remove}[1]{}

\newcommand{\IR}{\mathbb{R}}

\title{A linear time algorithm to cover and hit a set of
line segments optimally by two axis-parallel squares\footnote{A preliminary version of this paper
appeared in COCOON 2017, pages 457-468}}
\author{Sanjib Sadhu$^1$ \and Sasanka Roy$^2$
\and  Subhas C. Nandy$^2$ \and Suchismita Roy$^1$}
\date{$^1$Dept. of CSE, National Institute of Technology Durgapur,
India\\ $^2$Indian Statistical Institute, Kolkata, India}

\begin{document}
\maketitle
\begin{abstract}
This paper discusses the problem of covering and hitting 
a set of line segments $\cal L$  in $\IR^2$ by 
a pair of axis-parallel squares such that the side length of the larger of 
the two squares is minimized. We also discuss the restricted version of 
covering, where each line segment in $\cal L$ is to be covered completely 
by at least one square. The proposed algorithm for the covering problem 
reports the optimum result by executing only two passes of reading the 
input data sequentially. The algorithm proposed for the hitting and restricted 
covering problems produces optimum result in $O(n)$ time. All the proposed 
algorithms are in-place, and they use only $O(1)$ extra space. 
The solution of these problems also give a $\sqrt{2}$ 
approximation for covering and hitting those line segments $\cal L$
by two congruent disks of minimum radius with same computational complexity. 

 \end{abstract}
 
\noindent \textbf{Keywords.} Two-center problem, covering line segments by squares,
two pass algorithm, computational geometry 

\renewcommand*{\thefootnote}{\fnsymbol{footnote}}

\section{Introduction}
Covering a point set by squares/disks has drawn interest
to the researchers due to its applications in sensor network. 
Covering a given point set by $k$ congruent disks of minimum radius, 
known as $k$-center problem, is NP-Hard~\cite{marchetti}. For $k=2$, 
this problem is referred to as the 
{\em two center problem}~\cite{timothy,david,john,JaromczykK94,KatzKS00,Sharir97}. 

A line segment $\ell_i$ is said to be covered (resp. hit) by two squares
if every point (resp. at least one point) of $\ell_i$ lies inside one or 
both of the squares. For a given set $\cal L$ of line segments, the 
objective is 
to find two axis-parallel congruent squares such that each line segment 
in  $\cal L$ is covered (resp. hit) by the union of these two squares, and 
the size of the squares is as small as possible. There are mainly 
two variations of the covering problem: standard version and 
discrete version. In discrete version, the center of the squares 
must be on some specified points, whereas there are 
no such restriction in standard version. In this paper, we focus our 
study on the standard version of covering and hitting a set $\cal L$ of line 
segments in $\IR^2$ by two axis-parallel congruent squares of minimum size. 

As an application, consider a sensor network, where 
each mobile sensor is moving to and fro along different line segment. 
The objective is to place two base stations of minimum transmission 
range so that each of mobile sensors are always (resp. intermittently) 
connected to any of the base stations. This problem is exactly same as 
to cover (resp. hit) the line segments by two congruent disks (in our case 
axis-parallel congruent squares) of minimum radius.

Most of the works on the {\em two center problem} deal with covering 
a given point set. Kim and Shin~\cite{kim} provided an optimal 
solution for the {\em two center problem} of a convex polygon where 
the covering objects are two disks. As mentioned in \cite{kim}, the 
major differences between the {\it two-center problem for a convex 
polygon $P$} and the {\it two-center problem for a point set $S$} 
are (i) points covered by the two disks in the former problem are 
{\it in convex positions} (instead of arbitrary positions), and (ii) 
the union of two disks should also cover the edges of the polygon 
$P$. The feature (i) indicates the problem may be easier than the 
standard two-center problem for points, but feature (ii) says that 
it might be more difficult. To the best of 
our knowledge, there are no works on covering or hitting a set of line segments
 by two congruent squares of minimum size.

{\bf Related Work: }
Drenzer~\cite{journals/Drezner95} covered a given point set $S$
by two axis-parallel squares of minimum size in $O(n)$ time,
where where $n=|S|$.
 Ahn and Bae~\cite{journals/ijcga/Kim} proposed an 
$O(n^2\log n)$ time algorithm for covering a given point set $S$ 
by two disjoint rectangles where one of the rectangles is axis 
parallel and other one is of arbitrary orientation, and the area 
of the larger rectangle is minimized. Two congruent squares of minimum 
size covering all the points in $S$, where each one is of arbitrary 
orientation, can be computed in $O(n^4\log n)$ time 
\cite{conf/cocoa/Bhattacharya}. The best known deterministic 
algorithm for the standard version of two-center problem for a 
point set $S$ is given by Sharir~\cite{Sharir97} that runs in 
$O(n\log^9n)$ time. Eppstein~\cite{david} proposed a randomized 
algorithm for the same problem with expected time complexity 
$O(n\log^2 n)$. The standard and discrete versions of the 
two-center problem for a convex polygon $P$ was first solved by 
Kim and Shin~\cite{kim} in $O(n\log^3n\log\log n)$ and 
$O(n\log^2n)$ time respectively. Hoffmann~\cite{journals/comgeo/Hoffmann05}
solved the rectilinear 3-center problem for a point set in $O(n)$ time.
However none of the algorithms 
in~\cite{conf/cocoa/Bhattacharya,journals/Drezner95,journals/comgeo/Hoffmann05} 
can handle the line segments.

{\bf Our Work: } We propose  in-place algorithms for covering and hitting $n$
line segments in $\IR^2$ by two axis-parallel congruent squares of 
minimum size. We also study the restricted version of the covering problem 
where each object needs to be completely covered by at least one of the 
reported squares. The time complexities of our proposed algorithms for these 
three problems are $O(n)$, and 
these work using $O(1)$ extra work-space. The same algorithms work for 
covering/hitting a polygon, or a set of polygons by two axis-parallel 
congruent squares of minimum size. We show that the result of this algorithm 
can produce a solution for the problem of covering/ hitting these line 
segments by two congruent disks of minimum radius in $O(n)$ 
time  with an approximation factor $\sqrt{2}$.

\subsection{Notations and terminologies used } 
Throughout this paper, unless otherwise stated a {\em square} is used to imply
an axis-parallel square. We will use the following notations and definition.

\begin{tabular}{|c|l|}
\hline
{~\bf Symbols used~} & \hspace{2cm} {\bf Meaning} \\ \hline
$\overline{pq}$ and $|{pq}|$ & ~the line segment joining two points $p$ and $q$, 
and its length \\ \hline
$x(p)$ (resp. $y(p)$) & ~$x$- (resp. $y$-) coordinates of the point $p$  \\ \hline
$~|x(p)-x(q)|~$ & ~{\it horizontal distance} between a pair of points $p$ and $q$ \\ \hline
 $~|y(p)-y(q)|~$ & ~{\em vertical distance} between a pair of points $p$ and $q$ \\ \hline
 $s\in \overline{pq}$ & ~the point $s$ lies on the line segment $\overline{pq}$ \\ \hline
 $\Box efgh$ & ~an axis-parallel rectangle with vertices at $e$, $f$, $g$ and $h$\\ \hline
  {\em size($\cal S$)} &~size of square $\cal S$; it is the length of its one side \\ \hline
{\em $LS({\cal S})$, $RS({\cal S})$} &~Left-side of square $\cal S$ and right-side of square $\cal S$\\ \hline
{\em $TS({\cal S})$, $BS({\cal S})$} &~Top-side of square $\cal S$ and bottom-side of square $\cal S$\\ \hline
\end{tabular}

\begin{definition} 
A square is said to be {\bf anchored} 
with a vertex of a rectangle ${\cal R}=\Box efgh$, if one of
the corners of the square coincides with that vertex of ${\cal R}$.
\end{definition}

\section{Covering line segments by two congruent squares} 

\begin{description}
\item[LCOVER problem:] Given a set ${\cal L}=
\{\ell_1,\ell_2,\ldots,\ell_n\}$ of $n$ line segments (possibly intersecting) 
in $\IR^2$, the objective is to compute two  congruent 
squares ${\cal S}_1$ and ${\cal S}_2$ of minimum size whose union covers all the members in $\cal L$.
\end{description}

\begin{figure}[htbp]
\begin{minipage}[b]{0.5\linewidth}
\centering
\includegraphics[width=1\textwidth]{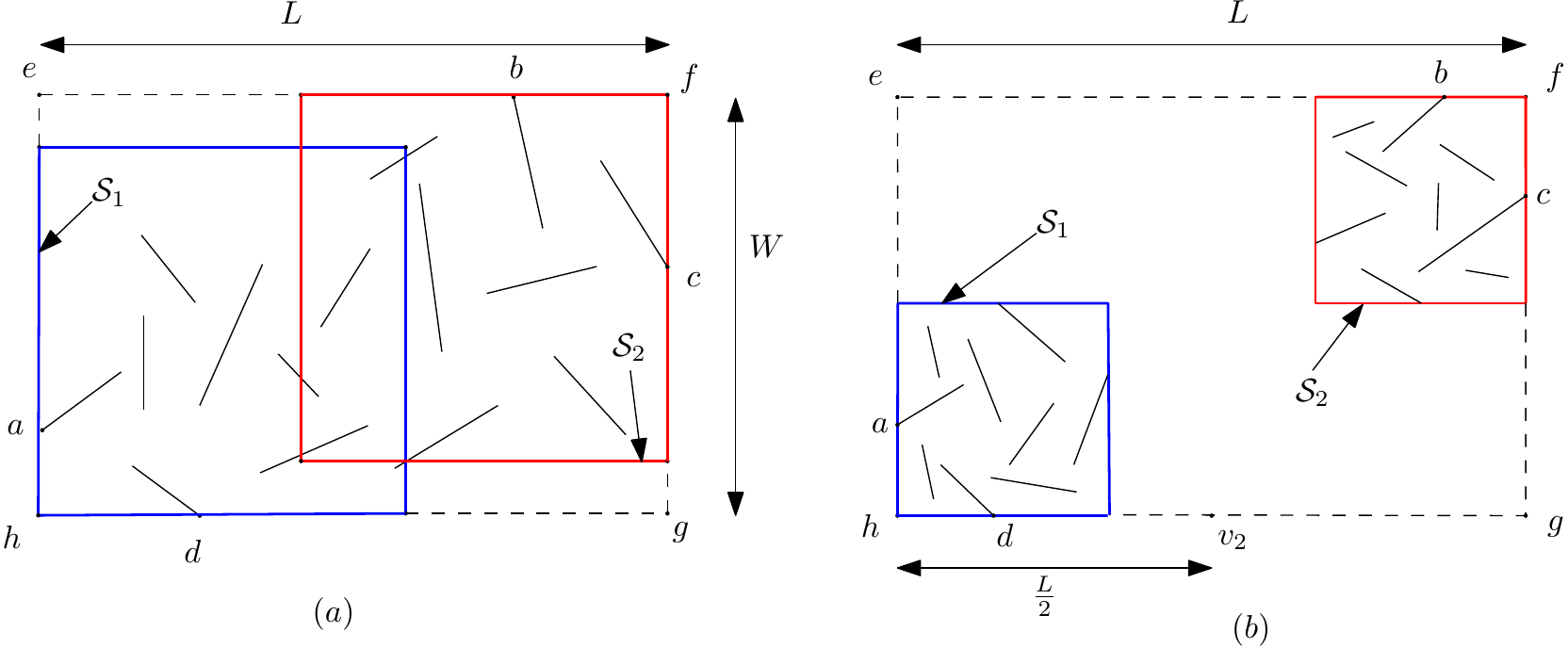}
\caption{Squares ${\cal S}_1$ and ${\cal S}_2$ are (a) overlapping, (b) disjoint.}
\label{fig5}
\end{minipage} %
\hspace{0.1cm}
\begin{minipage}[b]{0.5\linewidth}
\centering
\includegraphics[width=1\textwidth]{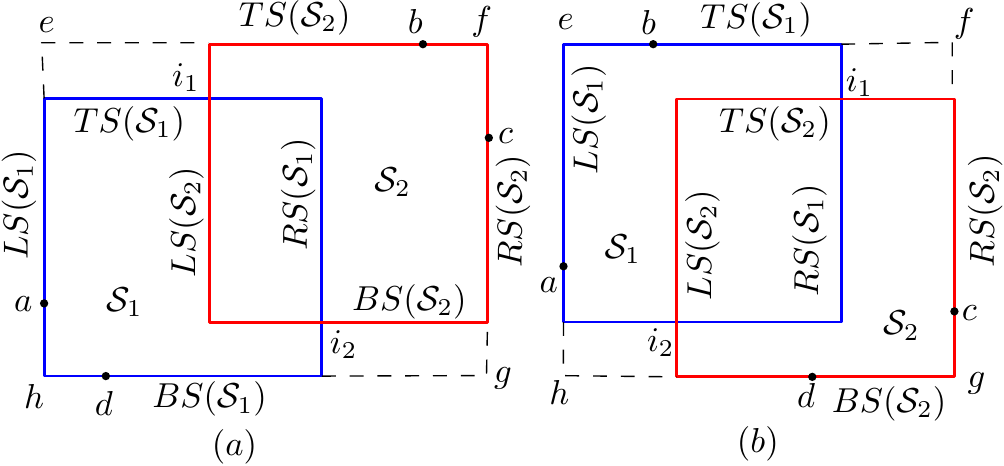}
 \caption{(a) {\bf Configuration~1} and (b) {\bf Configuration~2} of squares ${\cal S}_1$
 and ${\cal S}_2$.}
\label{general}
 \end{minipage} %
\end{figure}

In the first pass, a linear scan is performed among the objects in $\cal L$, 
and four points $a$, $b$, $c$ and $d$ are identified with minimum $x$-, maximum 
$y$-, maximum $x$- and minimum $y$-coordinate respectively among the end-points 
of $\cal L$. This defines an axis-parallel rectangle ${\cal R}=\Box efgh$ of 
minimum size that covers $\cal L$, where $a\in\overline{he}$, $b\in\overline{ef}$, 
$c\in \overline{fg}$ and $d \in \overline{gh}$.
 We use $L=|x(c)-x(a)|$ and $W=|y(b)-y(d)|$ as 
the length and width respectively of the rectangle ${\cal R} = 
\Box efgh$, and we assume 
that $L\geq W$.  We  assume that ${\cal S}_1$ 
lies to the left of ${\cal S}_2$. ${\cal S}_1$ and ${\cal S}_2$ may 
or may not overlap (see Fig.~\ref{fig5}). We use $\sigma=size({\cal S}_1)=size({\cal S}_2)$.

\begin{lemma}
\label{fact1} (a)
 There exists an optimal solution of the problem where
 $LS({\cal S}_1)$ and $RS({\cal S}_2)$ pass through the points $a$ 
 and $c$ respectively.

 (b) The top side of at least one of ${\cal S}_1$ and ${\cal S}_2$ pass through 
 the point $b$, and the bottom side of at least one of ${\cal S}_1$ and ${\cal S}_2$ 
 pass through the point $d$.  
 \end{lemma}
\remove{
\begin{proof} 

Follows from the optimality of the size of ${\cal S}_1$ and ${\cal S}_2$
for covering the segments in $\cal L$. \qed
\end{proof}
}

Thus in  an optimal solution of the {\bf LCOVER} problem, 
$a\in LS({\cal S}_1)$ and $c\in RS({\cal S}_2)$.
We need to consider two possible 
configurations of an optimum solution (i) $b\in TS({\cal S}_2)$ and 
$d\in BS({\cal S}_1)$, and (ii) $b\in TS({\cal S}_1)$ and $d\in BS({\cal S}_2)$. 
These are named as {\bf Configuration~1} and {\bf Configuration~2} respectively 
(see Fig.~\ref{general}).  

\begin{observation} \label{ob}
(a) If the optimal solution of LCOVER problem satisfies {\bf Configuration~1}, then 
the bottom-left corner of ${\cal S}_1$ will be anchored at the point $h$, and the top-right corner of 
${\cal S}_2$ will be anchored at the point $f$. 

(b) If the optimal solution of LCOVER problem satisfies {\bf Configuration~2}, then 
the top-left corner of ${\cal S}_1$ will be anchored at the point $e$, and the bottom-right corner of 
${\cal S}_2$ will be anchored at the point $g$. 
\end{observation}

We consider each of the configurations separately, and compute the two axis-parallel congruent squares 
${\cal S}_1$ and ${\cal S}_2$ of minimum size whose union covers the given 
set of line segments $\cal L$. If $\sigma_1$ and $\sigma_2$ are 
the sizes obtained for {\bf Configuration~1} and {\bf Configuration~2} respectively,  
then we report $\min(\sigma_1,\sigma_2)$.

\begin{figure}[t]
\begin{minipage}[b]{0.5\linewidth}
\centering
\includegraphics[width=0.8\textwidth]{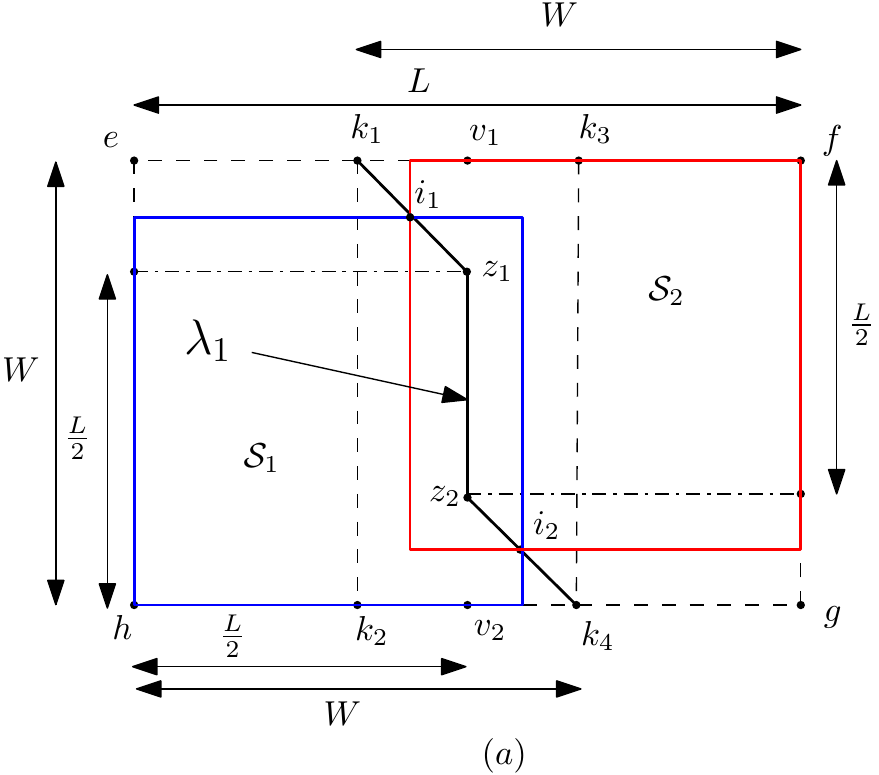}
\end{minipage} 
\hspace{0.3cm}
\begin{minipage}[b]{0.5\linewidth}
\centering
\includegraphics[width=0.8\textwidth]{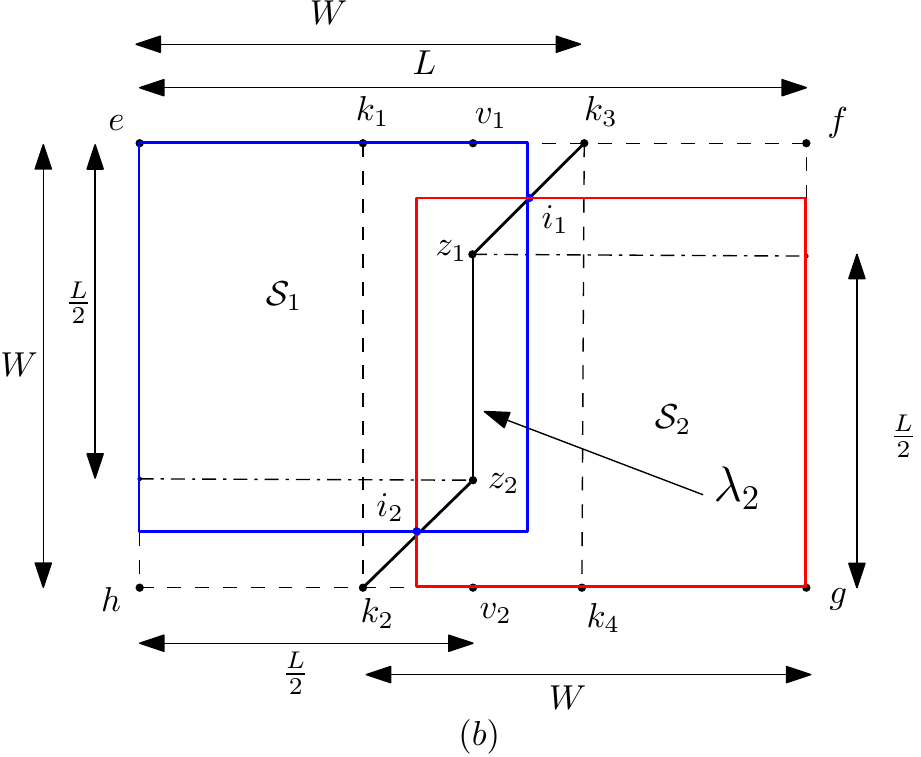}
\end{minipage}  
\caption{Voronoi partitioning line (a) $\lambda_1=k_1z_1z_2k_4$ of $f$ 
and $h$ in {\bf Configuration~1} 
(b) $\lambda_2=k_3z_1z_2k_2$ of $e$ and $g$ in {\bf Configuration~2}}
\label{simple_polygon_cover}
\end{figure}

 Consider
 the rectangle ${\cal R}=\Box efgh$ covering $\cal L$, and take six points
 $k_1$, $k_2$, $k_3$, $k_4$, $v_1$ and $v_2$ on the boundary of $\cal R$ satisfying 
 $|k_1f|=|ek_3|=|hk_4|=|k_2g|=W$ and
$|ev_1|=|hv_2|=\frac{L}{2}$ (see Fig.~\ref{simple_polygon_cover}).
Throughout the paper we assume $h$ as the origin in the co-ordinate system, i.e.
 $h=(0,0)$.
 

\begin{observation}
\label{vor_separator}
\begin{description}
\item[(i)] The Voronoi partitioning line $\lambda_1$ of the corners $f$ and $h$ of 
${\cal R}=\Box efgh$ with respect to the $L_\infty$ 
norm\footnote{$L_\infty$ distance between two points $a$ and $b$ is 
given by $\max(|x(a)-x(b)|,|y(a)-y(b)|)$} is the polyline  $k_1z_1z_2k_4$,
where the coordinates of its defining points are $k_1=(L-W,W)$, $z_1=(L/2,L/2)$, 
$z_2=(L/2,W-L/2)$ and $k_4=(W,0)$ (see Fig.~\ref{simple_polygon_cover}(a)).
\item[(ii)]  The Voronoi partitioning line $\lambda_2$ of $e$ and $g$ of ${\cal R}=\Box efgh$
in $L_\infty$ norm is the polyline  $k_3z_1z_2k_2$  where $k_3=(W,W)$ 
and $k_2=(L-W,0)$ (see Fig.~\ref{simple_polygon_cover}(b)).
\end{description}

Note that, if $W\leq \frac{L}{2}$, then the voronoi partitioning 
lines $\lambda_1$ and $\lambda_2$  for both the pairs
($f,~h$) and ($e,~g$)  will be same, i.e., 
$\lambda_1=\lambda_2=\overline{v_1v_2}$, where $v_1=(\frac{L}{2},0)$
and $v_2=(\frac{L}{2},W)$. 
\end{observation}

\begin{lemma} 
\label{voronoi}
(a) For {\bf Configuration~1}, 
All the points $p$ inside the polygonal region $ek_1z_1z_2k_4h$ satisfy
$d_\infty(p,h) < d_\infty(p,f)$, and all points $p$  
 inside the polygonal region $k_1fgk_4z_2z_1$ satisfy $d_\infty(p,f) < 
d_\infty(p,h)$ (see Fig.~\ref{simple_polygon_cover}(a)).

(b) Similarly for {\bf Configuration~2}, all points $p$ inside
polygonal region $ek_3z_1z_2k_2h$, satisfy $d_\infty(p,e) < d_\infty(p,g)$, and  
 all points $p$ that lie inside 
the polygonal region $k_3fgk_2z_2z_1$, satisfy $d_\infty(p,g) < d_\infty(p,e)$ 
(see Fig.~\ref{simple_polygon_cover}(b)). 
   \end{lemma}

\begin{lemma} \label{lmx}
\label{i_1_i_2}
If ${\cal S}_1$ and ${\cal S}_2$ intersect, then the points of intersection 
$i_1$ and $i_2$ will always lie on voronoi partitioning line  $\lambda_1 = 
k_1z_1z_2k_4$ (resp. $\lambda_2 = k_3z_1z_2k_2$) depending on whether 
${\cal S}_1$ and ${\cal S}_2$ satisfy {\bf Configuration~1} or {\bf 
Configuration~2}.
\end{lemma}
\remove{
\begin{proof}
 In {\bf Configuration~1} (resp.{\bf Configuration~2}), the result follows from the 
 equality of the distance of the intersecting points $i_1$ and $i_2$ from
 $f$ and $h$ (resp. $e$ and $g$), respectively.
 \qed
\end{proof}
}

Our algorithm 
consists of two passes. In each pass we  sequentially read each element of 
the input array $\cal L$ exactly once. We 
consider $W>\frac{L}{2}$ only.
The other case i.e $W\leq \frac{L}{2}$ can be handled in
the similar way.

  {\bf Pass-1 :} We compute the rectangle ${\cal R}= \Box efgh$, and the voronoi 
  partitioning lines $\lambda_1$ and $\lambda_2$ (see Fig. \ref{simple_polygon_cover}) for handling {\bf Configuration~1} and {\bf Configuration~2}.
  We 
  now discuss Pass 2 for {\bf Configuration~1}. The same method works for {\bf 
  Configuration~2}, and for both the configurations, the execution run simultaneously
   keeping a $O(1)$ working storage.

  {\bf Pass-2 :} $\lambda_1$ splits $\cal R$ into two disjoint parts, 
  namely ${\cal R}_1=\text{region}~ek_1z_1z_2k_4h$ and ${\cal R}_2=
  \text{region}~fk_1z_1z_2k_4g$. We  
  initialize $\sigma_1=0$. Next, we read elements in the input array ${\cal L}$ 
  in sequential manner. For each element $\ell_i=[p_i,q_i]$, we identify its 
  portion lying in one/both the parts ${\cal R}_1$ and ${\cal R}_2$. Now, 
  considering Lemma \ref{voronoi} and Observation \ref{ob}, we execute the following:
  \begin{description}
   \item[$\ell_i$ lies inside ${\cal R}_1$:] Compute $\delta=\max(d_\infty(p_i,h), 
   d_\infty(q_i,h))$.
   \item[$\ell_i$ lies inside ${\cal R}_2$:] Compute $\delta=\max(d_\infty(p_i,f), 
   d_\infty(q_i,f))$.
   \item[$\ell_i$ is intersected by $\lambda_1$:] Let $\theta$ be the point of intersection 
   of $\ell_i$ and $\lambda_1$, $p_i \in {\cal R}_1$ and $q_i \in {\cal R}_2$.
   Here, we compute $\delta=\max(d_\infty(p_i,h), d_\infty(\theta,h), d_\infty(q_i,f))$.
  \end{description}
If $\delta > \sigma_1$, we update $\sigma_1$ with $\delta$. Similarly, $\sigma_2$ 
is also computed in this pass considering the pair($e,~g$) and their 
partitioning line $\lambda_2$. 
Finally,  $\min(\sigma_1,\sigma_2)$ is returned as the optimal size along with 
the centers of the squares ${\cal S}_1$ and ${\cal S}_2$.  

\begin{theorem}
\label{convex_polygon}
Given a set of line segments $\cal L$ in $\IR^2$ in an array, one can 
compute two axis-parallel congruent squares of minimum size whose union covers 
$\cal L$ by reading the input array only twice in sequential manner, and 
maintaining $O(1)$ extra work-space. 
\end{theorem}
\remove{
\begin{proof}
The correctness of the algorithm follows from the facts that (i) we have only two 
configurations of the optimum solution (see Observation \ref{ob}), (ii) in 
Configuration 1, for every point $\theta$ in the left-partition (resp. right-partition) 
$d_\infty(\theta,h)$ $<$ (resp. $>$) $d_\infty(\theta,f)$ (similar things in Configuration 2), and (iii) 
we are covering portions of the members in $\cal L$ in the left (resp. right) partition by ${\cal S}_1$
(resp. ${\cal S}_2$).

As mentioned in the algorithm, we scanned the input array only twice in 
sequential manner. For each input element, we computed $\theta$ for both 
{\bf Configuration-1} and {\bf Configuration-2} in $O(1)$ time, and updated 
$\sigma_1$ and $\sigma_2$ if needed. The extra space required is to store 
all the variables $e$, $f$, $g$, $h$, $\lambda_1$, $\lambda_2$, $\sigma_1$, 
$\sigma_2$, $\delta$ is $O(1)$. 
\end{proof}
}
 
\section{Hitting line segments by two congruent squares} 
\begin{definition} 
\label{hit}
A geometric object $Q$ is said to be {\em hit} by a square ${\cal S}$
 if at least one point of $Q$ lies inside 
(or on the boundary of) ${\cal S}$.
\end{definition}
\begin{description}
\item[Line segment hitting (LHIT) problem:] Given a set ${\cal L}=
\{\ell_1,\ell_2,\ldots,\ell_n\}$ of $n$ line segments 
in $\IR^2$, compute two axis-parallel congruent 
squares ${\cal S}_1$ and ${\cal S}_2$ of minimum size whose union hits all the line segments in $\cal L$.
\end{description}

\begin{figure}[htbp] \vspace{-0.2in}
\centering
\includegraphics[width=0.7\textwidth]{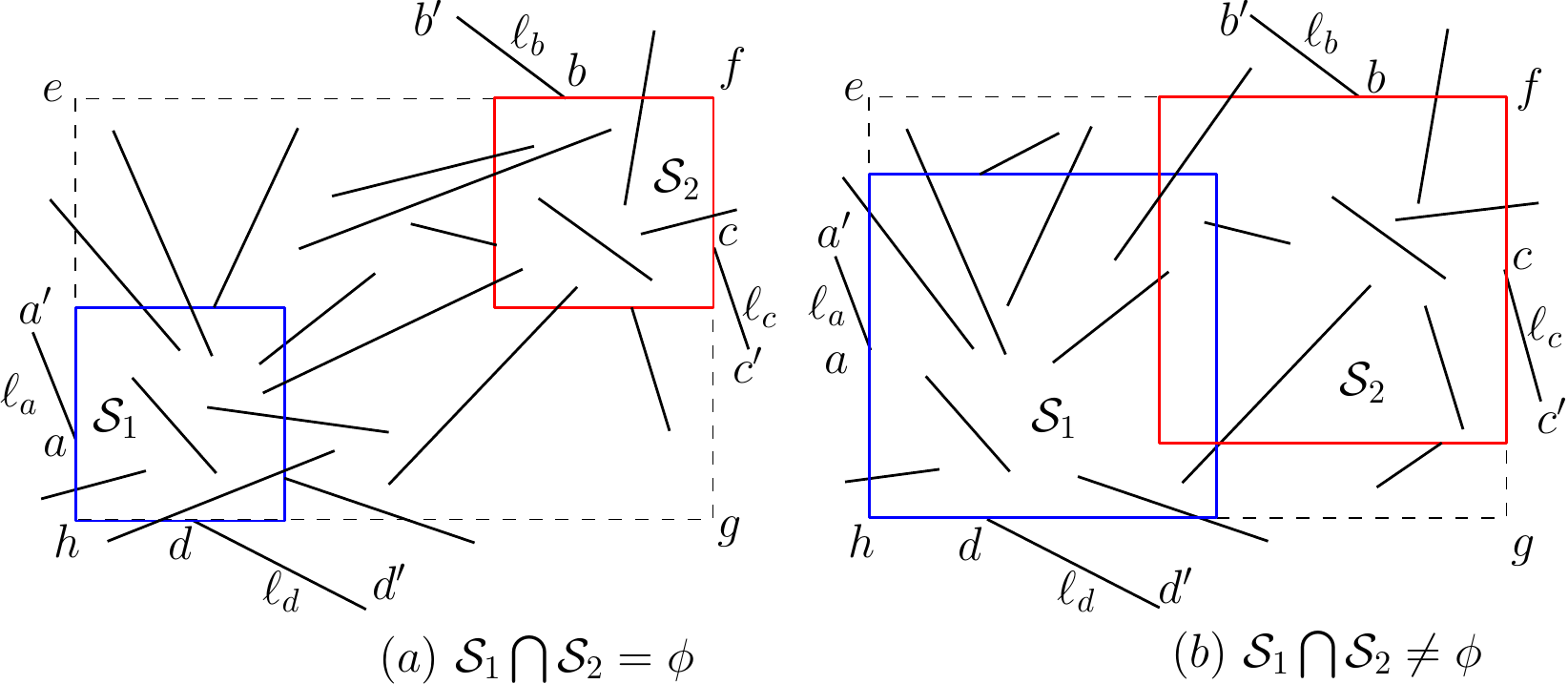}
 \caption{Two axis-parallel congruent squares ${\cal S}_1$
and ${\cal S}_2$ hit line segments in $\cal L$} \vspace{-0.1in}
\label{fig_p1} 
\end{figure}

The squares ${\cal S}_1$ and 
${\cal S}_2$ may or may not be disjoint (see Fig.
\ref{fig_p1}).
We now describe the algorithm for this {\bf LHIT} problem. 

For each line segment $\ell_i$,
we use $LP(\ell_i)$, $RP(\ell_i)$, $TP(\ell_i)$ and $BP(\ell_i)$
to denote its left end-point,
right end-point, top end-point and 
bottom end-point using the relations  
$x(LP(\ell_i))\leq x(RP(\ell_i))$
and $y(BP(\ell_i))\leq y(TP(\ell_i))$.
Now we compute 
four line segments $\ell_a,~\ell_b,~\ell_c,\text{and}~\ell_d\in {\cal L}$
such that 
one of their end-points $a$, $b$, $c$ and $d$, respectively
satisfy the following 
\[a=\min\limits_{\forall \ell_i\in {\cal L}} x(RP(\ell_i)),
b=\max\limits_{\forall \ell_i\in {\cal L}} y(BP(\ell_i)),
c=\max\limits_{\forall \ell_i\in {\cal L}} x(LP(\ell_i)),
d=\min\limits_{\forall \ell_i\in {\cal L}} y(TP(\ell_i))\]

We denote the other end point of $\ell_a$, $\ell_b$, $\ell_c$
and $\ell_d$ by $a'$, $b'$, $c'$ and $d'$, respectively.
 The four  points
$a$, $b$, $c$, $d$  define an axis-parallel rectangle ${\cal R}=\Box efgh$ of 
minimum size that hits all the members of $\cal L$ (as per Definition~\ref{hit}),
where $a\in\overline{he}$, $b\in\overline{ef}$, $c\in \overline{fg}$ and $d \in 
\overline{gh}$ (see Fig.~\ref{fig_p1}).  
We use $L=|x(c)-x(a)|$ and $W=|y(b)-y(d)|$ as the length and width 
of the rectangle ${\cal R}$, and assume $L \geq W$. 
Let ${\cal S}_1$ and ${\cal S}_2$ be the 
two axis-parallel congruent squares that hit the given 
line segments $\cal L$
optimally, where 
${\cal S}_1$ lies to the left of ${\cal S}_2$.

\begin{observation}
\label{fact_obs1} (a)
 The left side of ${\cal S}_1$ (resp. right side of ${\cal S}_2$) must not 
 lie to the right of (resp. left of) the point $a$ 
 (resp. $c$), and 
 (b) the top side (resp. bottom side) of both ${\cal S}_1$ 
 and ${\cal S}_2$ cannot lie below (resp. above) 
 the point $b$ (resp. $d$).
 \end{observation}
 
 For the {\bf LHIT} problem, we say 
 ${\cal S}_1$ and ${\cal S}_2$ are in {\bf Configuration~1},
 if ${\cal S}_1$ 
 hits both $\ell_a$ and $\ell_d$, and ${\cal S}_2$ hits both  $\ell_b$ and $\ell_c$.
 Similarly,
 ${\cal S}_1$ and ${\cal S}_2$ are said to be in {\bf Configuration~2},
 if ${\cal S}_1$ hits both $\ell_a$ and $\ell_b$, and ${\cal S}_2$
 hits both $\ell_c$ and $\ell_d$. 
 
 Without loss of generality, we assume that ${\cal S}_1$ and ${\cal S}_2$
 are in {\bf Configuration~1}.
 We compute the reference (poly) line ${\mathbb D}_1$ 
 (resp. ${\mathbb D}_2$) on which the
 top-right corner of ${\cal S}_1$ (resp. bottom-left corner of ${\cal S}_2$) will lie.
 Let ${\mathbb T}_1$ (resp. ${\mathbb T}_2$) be the line
 passing through $h$ (resp. $f$) with slope $1$. 
 Our algorithm consists of the following phases: 
 
 \begin{description}
  \item[1] Computation of the reference lines ${\mathbb D}_1$ and ${\mathbb D}_2$.
  \item[2] For each line segment $\ell_i\in {\cal L}$, computation of
  the size of the minimum square ${\cal S}_1$ (resp. ${\cal S}_2$) required to hit
$\ell_i$, $\ell_a$ and $\ell_d$ (resp. $\ell_i$, $\ell_b$ and $\ell_c$), 
where the top-right (resp. bottom-left) corner of ${\cal S}_1$ (resp. ${\cal S}_2$) 
lies on $\mathbb{D}_1$ (resp. $\mathbb{D}_2$).
\item[3] Determining the pair (${\cal S}_1$, ${\cal S}_2$) that hit all the line segments in $\cal L$
and $\max(size({\cal S}_1),~\newline size({\cal S}_2))$ is minimized.
  \end{description}
  
 {\bf Computation of the reference lines ${\mathbb D}_1$ and ${\mathbb D}_2$}:
 The reference line ${\mathbb D}_1$ is computed based on the following
 four possible orientations of $\ell_a$ and $\ell_d$ 
\begin{figure}[t] 
\begin{minipage}[b]{0.5\linewidth}
\centering
\includegraphics[width=0.95\textwidth]{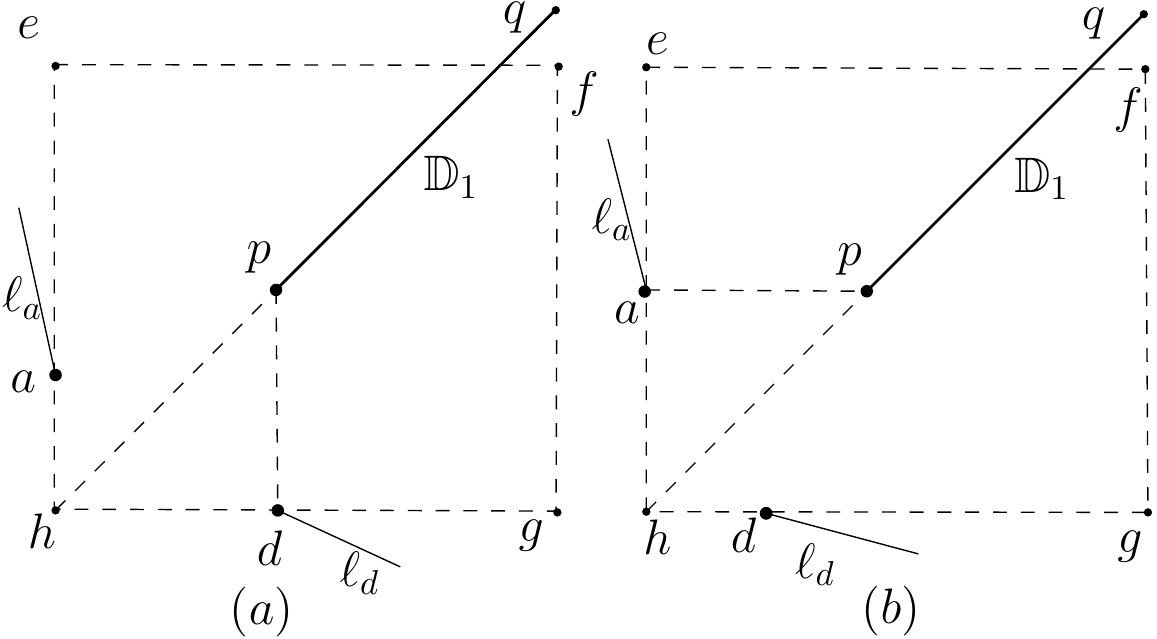}
\caption{${\mathbb D}_1$ for $y(LP(\ell_a))\geq y(RP(\ell_a))$ \newline
and $x(TP(\ell_d))< x(BP(\ell_d))$} 
\label{fig_ref1}
 \end{minipage} 
\hspace{0.1cm}
\begin{minipage}[b]{0.5\linewidth}
\centering
\includegraphics[width=0.95\textwidth]{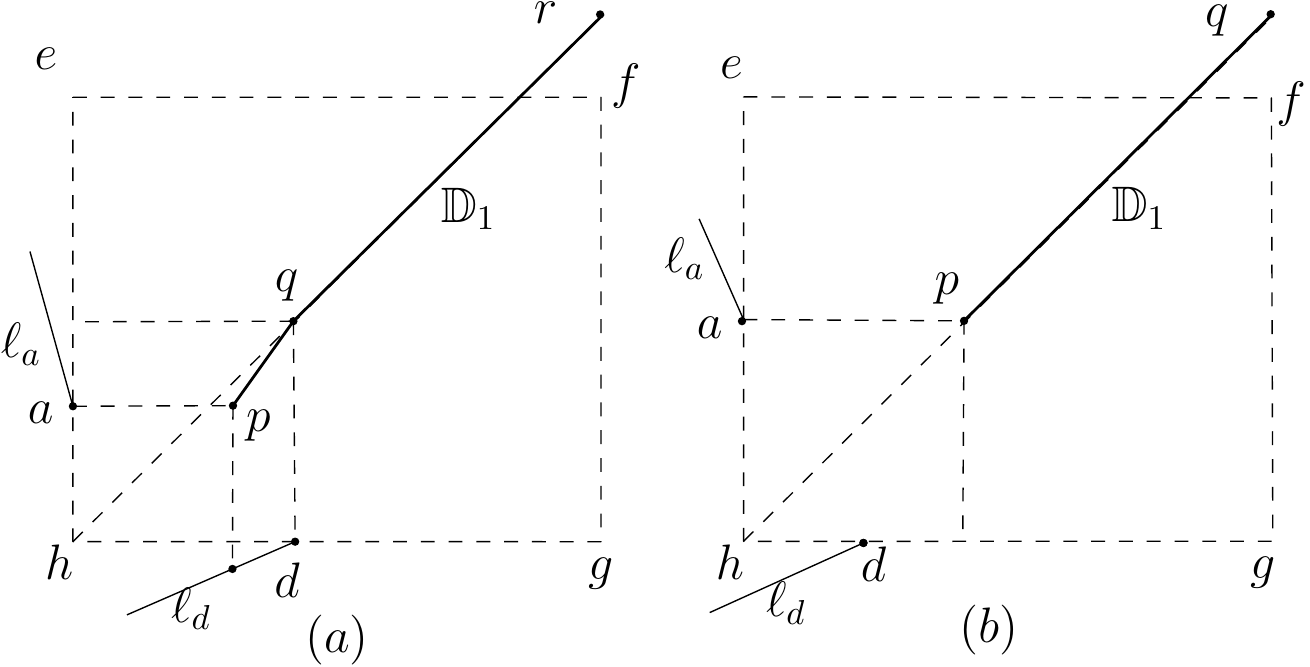}
\caption{${\mathbb D}_1$ for $y(LP(\ell_a))\geq y(RP(\ell_a))$ \newline and 
          $x(TP(\ell_d))\geq x(BP(\ell_d))$ } 
\label{fig_ref2}
\end{minipage} 
\end{figure}

\begin{figure}[htbp]
\begin{minipage}[b]{0.5\linewidth}
\centering
 \includegraphics[width=1\textwidth]{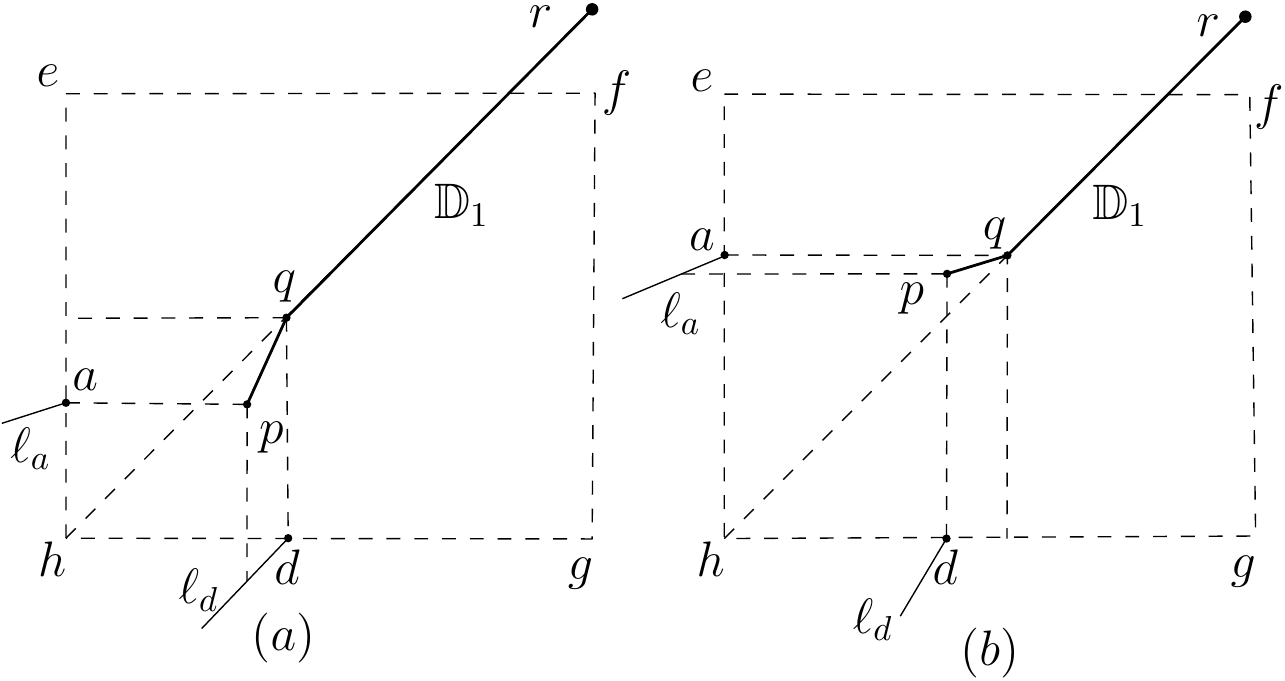}
\end{minipage} 
\begin{minipage}[b]{0.5\linewidth}
\centering
 \includegraphics[width=1\textwidth]{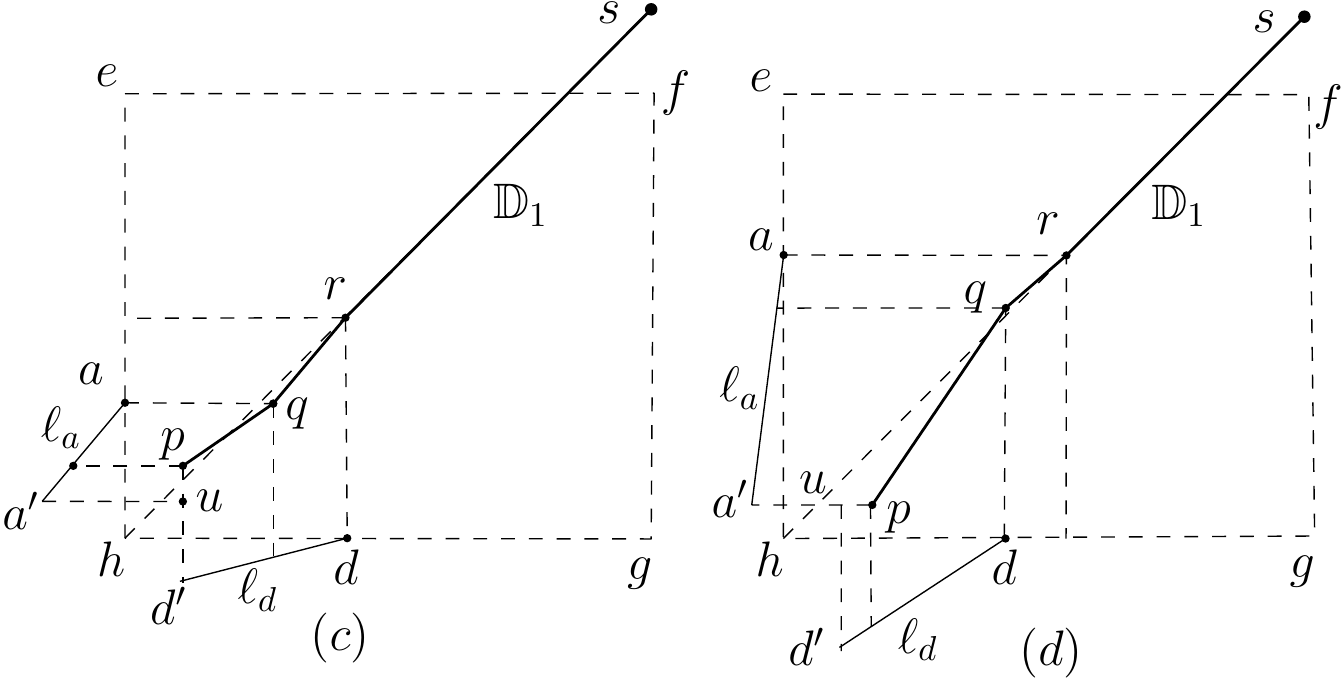}
\end{minipage}
 \caption{${\mathbb D}_1$ for $y(LP(\ell_a))<y(RP(\ell_a))$ and
      $x(TP(\ell_d))> x(BP(\ell_d))$ } 
\label{fig_ref4}
\end{figure}

 \begin{itemize}
  \item[(i)] {\boldmath  $y(LP(\ell_a))\geq y(RP(\ell_a))$} and
       {\boldmath $x(TP(\ell_d))< x(BP(\ell_d))$: }
       Here ${\mathbb D}_1$ is the segment $\overline{pq}$
       on ${\mathbb T}_1$ where $p$ is determined (i) by its $x$-coordinate
       i.e. $x(p)=x(d)$, if $|ha|<|hd|$ (see Fig.~\ref{fig_ref1}(a)),
       (ii) by its $y$-coordinate i.e. $y(p)=y(a)$, if $|ha|\geq|hd|$
       (see Fig.~\ref{fig_ref1}(b)).
       The point $q$ on ${\mathbb T}_1$ satisfy $x(q)=x(f)$.

  \item[(ii)] {\boldmath  $y(LP(\ell_a))\geq y(RP(\ell_a))$} and 
        {\boldmath $x(TP(\ell_d))\geq x(BP(\ell_d))$: } 
        Here, \\if $|ha|<|hd|$ (see Fig.~\ref{fig_ref2}(a)), then the reference line ${\mathbb D}_1$
        is a polyline $\overline{pqr}$, where (i) $y(p)=y(a)$
        and $x(p)$ satisfies $|x(p)-x(a)|=$ vertical distance of $p$
        from the line segment $\ell_d$, (ii) the point $q$ lies on ${\mathbb T}_1$
        satisfying $x(q)=x(d)$ and (iii) the point $r$ lies on ${\mathbb T}_1$ 
        satisfying $x(r)=x(f)$.\\
        If $|ha|\geq|hd|$ (see Fig.~\ref{fig_ref2}(b)), then the reference line ${\mathbb D}_1$ is a 
        line segment $\overline{pq}$, where $p$, $q$ lies on ${\mathbb T}_1$,
        and $p$ satisfies $y(p)=y(a)$ and $q$ satisfies $x(q)=x(f)$.

  \item[(iii)] {\boldmath  $y(LP(\ell_a))<y(RP(\ell_a))$} and
	    {\boldmath $x(TP(\ell_d))\leq x(BP(\ell_d))$: }
     This case is similar to case~$(ii)$, and we can compute the respective reference
    lines.
    \item[(iv)] {\boldmath  $y(LP(\ell_a))<y(RP(\ell_a))$} and
    {\boldmath $x(TP(\ell_d))> x(BP(\ell_d))$: } There are two possible subcases:
    \begin{itemize}
    \item[(A)] If $\ell_a$
    and $\ell_d$ are parallel or intersect (after extension)
    at a point to the right of $\overline{he}$ (Fig.~\ref{fig_ref4}(a,b)),
    then the reference line ${\mathbb D}_1$ is a polyline $\overline{pqr}$,
    where {\bf (a) if} $\mathbf{|ha|<|hd|}$ (Fig.~\ref{fig_ref4}(a)), then
    (1) $y(p)=y(a)$ and $|x(p)-x(a)|=$ the vertical distance of
	$p$ from $\ell_d$, (2) the points $q$ and $r$ lie on ${\mathbb T}_1$
        satisfying $x(q)=x(d)$ and $x(r)=x(f)$, 
        {\bf (b) if} $\mathbf{|ha|>|hd|}$ (Fig.~\ref{fig_ref4}(b)), 
        then (1) $x(p)=x(d)$ and $|y(p)-y(d)|=$ the horizontal distance of
	$p$ from $\ell_a$, (2) the points $q$ and $r$ lie on ${\mathbb T}_1$
        satisfying $y(q)=y(a)$ and $x(r)=x(f)$.\\
        
    \item[(B)] If extended $\ell_a$ and $\ell_d$ intersect at a point to the
    left of $\overline{he}$ (Fig.~\ref{fig_ref4}(c,d)), then 
    ${\mathbb D}_1$ is a polyline $\overline{pqrs}$, where
    
    (i) the line segment $\overline{pq}$ is such that for every point $\theta\in \overline{pq}$,
    the horizontal distance of $\theta$ from $\ell_a$ and the vertical distance of $\theta$
    from $\ell_d$ are same.
    
    (ii) the line segment $\overline{qr}$ is such that for every point $\theta\in \overline{qr}$,
    we have
    
      {\bf if} $\mathbf{|ha|<|hd|}$ then $|x(\theta)-x(a)|=$ vertical distance of $\theta$ from $\ell_d$
       (Fig.~\ref{fig_ref4}(c)), 
      {\bf else}  $|y(\theta)-x(d)|=$ horizontal distance of $\theta$ from $\ell_a$,
       (Fig.~\ref{fig_ref4}(d))
       
     (iii) the point $s$ lies on ${\mathbb T}_1$ 
        satisfying $x(s)=x(f)$.
 \end{itemize}
 \end{itemize}
In the same way, we can compute the reference line ${\mathbb D}_2$  based on the
 four possible orientations of $\ell_b$ and $\ell_c$. The break points/end points
 of ${\mathbb D}_2$ will be referred to as $p'$, $q'$, $r'$, $s'$
 depending on the appropriate cases. 
From now onwards, we state the position of square ${\cal S}_1$ (resp. ${\cal S}_2$)
in terms of the position of its top-right corner (resp. bottom-left corner).
\begin{observation}
\label{minimal}
 The point $p\in{\mathbb D}_1$ (resp. $p'\in{\mathbb D}_2$) gives
 the position of  minimum sized axis-parallel square ${\cal S}_1$
 (resp. ${\cal S}_2$)
 that hit $\ell_a$ and $\ell_d$ (resp. $\ell_b$ and $\ell_c$). 
\end{observation}

{\bf Computation of minimum
sized squares ${\cal S}_1$ and ${\cal S}_2$ to hit $\ell_i$ :}
Let $L_V$ (resp. $L_H$) denotes the vertical (resp. horizontal) half-line below
(resp. to the left of) the 
point $p\in {\mathbb D}_1$.
Similarly, $L'_V$ (resp. $L'_H$) denotes the vertical (resp. horizontal) half-line above
(resp. to the right of) the 
point $p'\in {\mathbb D}_2$.
Observe that, if a line segment  $\ell_i\in \cal L$ 
intersects with any of $L_H$ or $L_V$, or 
if $\ell_i$ lie completely below $L_H$ and to the left of $L_V$, then it ($\ell_i$) 
will be hit by any square that hits both $\ell_a$ and $\ell_d$. Similarly,  if a line segment $\ell_i$ 
intersects with any of $L'_H$ or $L'_V$; or if $\ell_i$
lies completely above $L'_H$ and to the right of $L'_V$, then it ($\ell_i$)
will be hit by any square that hits both $\ell_b$ and $\ell_c$. 
Thus, such line segments will not contribute any event point on ${\mathbb D}_1$ (resp. ${\mathbb D}_2$).

\begin{figure}[htbp]
\vspace{-0.1in}
\begin{minipage}[b]{0.5\linewidth}
\centering
\includegraphics[scale=0.77]{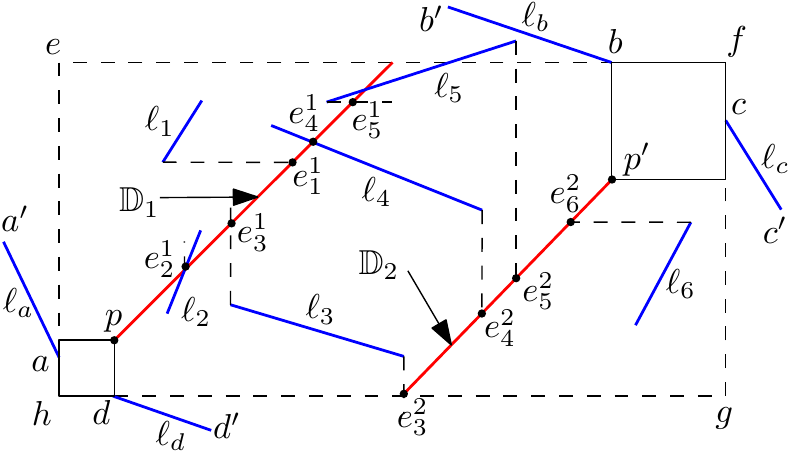}
\caption{Event points for {\bf LHIT} problem under {\bf Configuration~1}}
 \label{fig_p2}
\end{minipage} 
\hspace{0.1cm}
\begin{minipage}[b]{0.5\linewidth}
\centering
\includegraphics[scale=0.4]{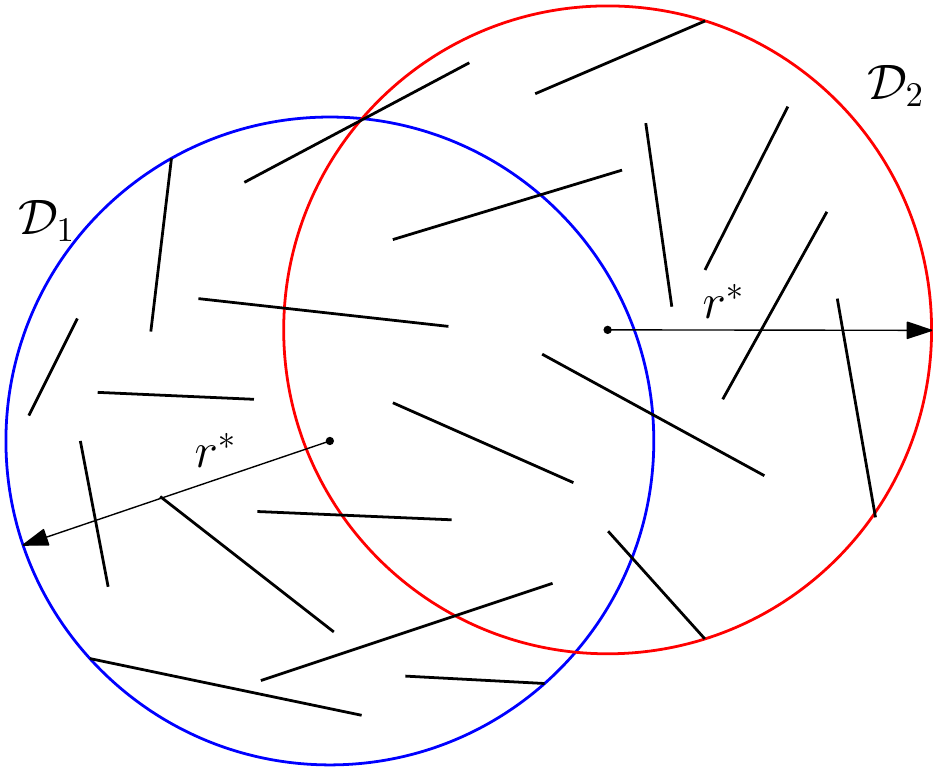}
 \caption{Covering $\cal L$ by two disks ${\cal D}_1$ $\&$ ${\cal D}_2$}
 \label{line_segmentx}
 \end{minipage} 
\end{figure}

For each of the line segments $\ell_i\in {\cal L}$, we create two event points 
$e_i^1 \in {\mathbb D}_1$ and $e_i^2 \in {\mathbb D}_2$, as follows:
\begin{description}
\item[(i)] If $\ell_i$ lies completely above ${\mathbb D}_1$ (resp. ${\mathbb D}_2$), then
we compute the event point $e_i^1=(x_{i_1},y_{i_1})$ on ${\mathbb D}_1$
(resp. $e_i^2=(x_{i_2},y_{i_2})$ on ${\mathbb D}_2$)
satisfying $y_{i_1}=y(BP(\ell_i))$
(resp. $x_{i_2}=x(RP(\ell_i))$).
 (see the points $e_1^1$ for $\ell_1$
and $e_4^2$ for $\ell_4$ in Fig.~\ref{fig_p2}).
\item[(ii)] If $\ell_i$ lies completely below ${\mathbb D}_1$ (resp. ${\mathbb D}_2$),
we compute the event point $e_i^1=(x_{i_1},y_{i_1})$ on 
${\mathbb D}_1$ (resp. $e_i^2=(x_{i_2},y_{i_2})$ on ${\mathbb D}_2$) satisfying $x_{i_1}=x(LP(\ell_i))$
(resp. $y_{i_2}=y(TP(\ell_i))$).
(see $e_3^1$ for $\ell_3$
and $e_6^2$ for $\ell_6$ in Fig.~\ref{fig_p2}).
\item[(iii)] If $\ell_i$ intersects with ${\mathbb D}_1$ (resp. ${\mathbb D}_2$) 
at point $p_1$ (resp. $q_1$), then
we create the event point $e_i^1$ on ${\mathbb D}_1$ 
(resp. $e_i^2$ on ${\mathbb D}_2$) according to the following rule:
      \begin{description}
       \item[(a)] If the $x(BP(\ell_i))>x(p_1)$ (resp. $x(TP(\ell_i))<x(q_1)$), 
      then we take $p_1$ (resp. $q_1$) as the event point $e^i_1$ (resp. $e^i_2$).
      (see $e_4^1$ for $\ell_4$ in Fig.~\ref{fig_p2}).
      \item[(b)] If  $x(BP(\ell_i))<x(p_1)$ then 
      if $BP(\ell_i)$ lies below ${\mathbb D}_1$ then
      we consider the point of intersection by ${\mathbb D}_1$ with
      the vertical line passing
      through the $BP(\ell_i)$ as the point $e_i^1$ (see
      $e_2^1$ for $\ell_2$ in Fig.~\ref{fig_p2}), and \\
      if $BP(\ell_i)$ lies above ${\mathbb D}_1$ then
      we consider the point of intersection ${\mathbb D}_1$
      with the horizontal line passing
      through  $BP(\ell_i)$ as the event point $e_i^1$ (see 
      $e_5^1$ for $\ell_5$ in Fig.~\ref{fig_p2}).
      \item[(c)] If 
      $x(TP(\ell_i))>x(q_1)$ then 
      if $TP(\ell_i)$ lies above ${\mathbb D}_2$ then
      we consider the point of intersection by ${\mathbb D}_2$ with
      the vertical line passing
      through  $TP(\ell_i)$ as the event point $e_i^2$, and  
      if $TP(\ell_i)$ lies below ${\mathbb D}_2$ then we consider
      the point of intersection ${\mathbb D}_2$
      with the horizontal line passing through $TP(\ell_i)$ as the event point $e^2_i$.
      \end{description}
\end{description}

  \begin{observation}
   \label{D1_D2}
    (i) An event $e_i^1$ on ${\mathbb D}_1$ shows the position of the top-right
   corner of the minimum sized square ${\cal S}_1$ that hits $\ell_a$, $\ell_d$ and $\ell_i$, and  
   an event $e_i^2$ on ${\mathbb D}_2$ shows the position of the 
   bottom-left corner of the minimum sized square ${\cal S}_2$ that hits
   $\ell_b$, $\ell_c$ and $\ell_i$.

   (ii) The square ${\cal S}_1$ whose top-right corner is at $e_i^1$ on ${\mathbb D}_1$ hits all those line
   segments $\ell_j$ whose corresponding event points $e_j^1$ on ${\mathbb D}_1$ 
   satisfies $x(h)\leq x(e_j^1)\leq x(e_i^1)$. Similarly,  
   the square  ${\cal S}_2$ whose bottom-left corner is at 
  $e^i_2$ on ${\mathbb D}_2$ hits all those line segments $\ell_j$ whose corresponding event point $e_j^2$ on  
  ${\mathbb D}_2$ satisfies  $x(e_i^1)\leq x(e_j^1) \leq x(f)$.
   \end{observation}

Thus for each line segment $\ell_i\in {\cal L}$, we have two parameters  
$\sigma_{i_1}$ and $\sigma_{i_2}$, where $\sigma_{i_1}$ (resp. $\sigma_{i_2}$) 
denotes the size of the minimum square required to hit $\ell_i$, $\ell_a$ and $\ell_d$ (resp. 
$\ell_i$, $\ell_b$ and $\ell_c$). It is to be noted that $p$  and $p'$ are also the 
event points on ${\mathbb D}_1$ and ${\mathbb D}_2$,
respectively (see Observation~\ref{minimal}). We now compute two minimum sized squares ${\cal S}_1$ and 
${\cal S}_2$ to hit all the line segments in $\cal L$ as follows:

   Let $\sigma_{{\min}_1}$ (resp. $\sigma_{{\min}_2}$) denote the size of
  the minimum square ${\cal S}_1$ (resp. ${\cal S}_2$) required to hit the line segments $\ell_a$
  and $\ell_d$ (resp. $\ell_b$ and $\ell_c$).
  Initially we compute these $\sigma_{\min_1}$
  and $\sigma_{\min_2}$ which are determined by
  the position of the point $p$ and $p'$ lying on ${\mathbb D}_1$
  and ${\mathbb D}_2$, respectively.
   Then for each line segment $\ell_i\in \cal L$,
   we compute $\sigma_{i_1}$ and $\sigma_{i_2}$, and compare between them.
  Our objective is to reduce the size of the larger square; hence
  if $\sigma_{i_1}\leq \sigma_{i_2}$, then we choose the square ${\cal S}_1$
  to hit $\ell_i$, otherwise we choose square ${\cal S}_2$.
  If $\sigma_{i_1}\leq \sigma_{i_2}$, then we compare $\sigma_{i_1}$ with $\sigma_{\min_1}$.
  If $\sigma_{i_1} > \sigma_{\min_1}$, then we update $\sigma_{\min_1}$
  as $\sigma_{i_1}$, otherwise $\sigma_{\min_1}$ remains same.
  On the other hand, if $\sigma_{i_1}> \sigma_{i_2}$, we compare $\sigma_{i_2}$
  with $\sigma_{\min_2}$, and update  $\sigma_{\min_2}$
  as $\sigma_{i_2}$ only if  $\sigma_{\min_2}$ is less than
  $\sigma_{i_2}$.
    After all the line segments have been processed sequentially, the $\max(\sigma_{\min_1},
  \sigma_{\min_2})$ will give the
  minimum size of the congruent squares ${\cal S}_1$ and ${\cal S}_2$
  to hit all the line segments of $\cal L$ in
    {\bf Configuration~1}.    
    It is to be noted that while processing the line segments in $\cal L$ sequentially,
  for each line segment $\ell_i \in {\cal L}$, we need to generate the two event points $(e_i^1,e_i^2)$, compute $(\sigma_{i_1},\sigma_{i_2})$, and use it to update
  $(\sigma_{{\min}_1},\sigma_{{\min}_2})$, and use the same locations for processing the next line segment $\ell_j \in {\cal L}$.
Hence, the aforesaid steps can be executed in
  linear time using $O(1)$ space.

    Similarly, we can determine the  optimal size of the congruent squares ${\cal S}_1$ and ${\cal S}_2$ in
    {\bf Configuration~2}.    
    Finally we consider that configuration for which the size of the congruent
    squares is minimized.
    Thus we have the following result:
    \begin{theorem}
     \label{opt_lhit}
     The {\bf LHIT} problem can be solved optimally in $O(n)$ time using $O(1)$ extra work-space.
    \end{theorem}
 \section{Restricted version of LCOVER problem}
 In restricted version of the {\bf LCOVER} problem,
 each line segment in $\cal L$ is to be covered completely
 by atleast one of the two congruent axis-parallel squares ${\cal S}_1$
 and ${\cal S}_2$.
 We compute the axis-parallel rectangle ${\cal R}=\Box efgh$ passing through
  the four points $a$, $b$, $c$ and $d$
 as in our algorithm for {\bf LCOVER} problem.
 As in the {\bf LCOVER} problem, here also we have two possible configurations for optimal solution.
 Without loss of generality, we assume that ${\cal S}_1$
 and ${\cal S}_2$ satisfy {\bf Configuration~1}.
 We consider two reference lines ${\mathbb D}_1$ and ${\mathbb D}_2$,
 each with unit slope 
 that passes through $h$ and $f$, respectively. These reference
 lines ${\mathbb D}_1$ and ${\mathbb D}_2$ are the locus of the top-right 
 corner of ${\cal S}_1$ and bottom-left corner of
 ${\cal S}_2$, respectively. 
 For each line segment $\ell_i$, we create an event point 
$e_i^1=(x_{i_1},y_{i_1})$ on ${\mathbb D}_1$ 
(resp. $e_i^2=(x_{i_2},y_{i_2})$ on ${\mathbb D}_2$) as follows:
\begin{description}

\item(i) If $\ell_i$ lies completely above ${\mathbb D}_1$ (resp. ${\mathbb D}_2$), then
the event point $e_i^1$ on ${\mathbb D}_1$ 
(resp. $e_i^2$ on ${\mathbb D}_2$) 
will satisfy $y_{i_1}=y(TP(\ell_i))$ (resp. $x_{i_2}=x(LP(\ell_i))$).
 \item(ii) If $\ell_i$ lies completely below ${\mathbb D}_1$ (resp. ${\mathbb D}_2$) then 
the event point $e_i^1$ on ${\mathbb D}_1$ (resp. $e_i^2$ on ${\mathbb D}_2$)
will satisfy $x_{i_1}=x(RP(\ell_i))$
(resp. $y_{i_2}=y(BP(\ell_i))$).
\item(iii) If $\ell_i$ intersects with ${\mathbb D}_1$ then
we create the event point $e_i^1$ on ${\mathbb D}_1$ as follows:\\
Let the horizontal line through 
       $TP(\ell_i)$ intersect with ${\mathbb D}_1$ 
       at point $p$, and the vertical  line through 
       $BP(\ell_i)$  intersect with ${\mathbb D}_1$ 
       at point $q$. If  $x(p)>x(q)$,
       then we take $p$ (else $q$) as the  event point on 
       ${\mathbb D}_1$.
\item(iv) If $\ell_i$ intersects with ${\mathbb D}_2$, then
we create the event point $e_i^2$ on ${\mathbb D}_2$ as follows: \\
Let the vertical line through 
       $BP(\ell_i)$ intersect with ${\mathbb D}_2$
       at point $p$, and the horizontal line through 
       $TP(\ell_i)$ intersect with ${\mathbb D}_2$
       at point $q$. If  $x(p)>x(q)$,
       then we take $q$ (else $p$) as the  event point on ${\mathbb D}_2$.

\end{description}
 Observation similar to Observation~\ref{D1_D2} in {LHIT}
 problem also
 holds for this problem where ${\cal S}_1$ and ${\cal S}_2$ cover $\cal L$
 with restriction. Thus, here we can follow the same technique as in {LHIT} problem to obtain the following result:
    \begin{theorem}
     \label{opt_restrict}
     The restricted version of {\bf LCOVER} problem can be solved optimally 
     in $O(n)$ time using $O(1)$ extra work-space.
    \end{theorem}
\section{Covering/Hitting line segments by two congruent disks} \label{2-center}
In this section, we consider problems related to {\bf LCOVER},
{\bf LHIT} and {\bf restricted LCOVER} problem, called {\em two center 
problem}, where the objective is to cover, hit or restricted-cover the given line segments in $\cal L$
by two congruent 
disks so that their (common) radius is minimized. Fig.~\ref{line_segmentx} demonstrates 
a covering instance of this {\em two center 
problem}. Here, we first compute  two axis-parallel squares ${\cal S}_1$ and ${\cal S}_2$ 
whose union covers/ hits all the members of $\cal L$ optimally as described in the 
previous section. Then we  
report the circum-circles ${\cal D}_1$ and ${\cal D}_2$ of 
${\cal S}_1$ and ${\cal S}_2$ respectively as an approximate solution of the {\em 
two center problem}.

\begin{lemma}
\label{lower_bound_lm}
A lower bound for the optimal radius of  two center problem for $\cal L$ is the radius 
$r'$ of in-circle of the two congruent squares ${\cal S}_1$ and 
${\cal S}_2$ of minimum size that cover/ hit/ restricted-cover $\cal L$; i.e. $r'\leq r^*$.
\end{lemma}

 The radius $r$ of the circum-circle ${\cal D}_1$ and ${\cal D}_2$ of the squares ${\cal S}_1$
 and ${\cal S}_2$ is $\sqrt{2}$ times of the radius $r'$ of their in-circles.
 Lemma~\ref{lower_bound_lm} says that $r'\leq r^*$.
 Thus, we have 
 
\begin{theorem}
\label{srt_2_th}
 Algorithm  {\bf Two center} generates a $\sqrt{2}$ approximation result for 
  {\em LCOVER}, {\em LHIT} and {\em restricted LCOVER}  problems for the line segments in $\cal L$.
\end{theorem}

\end{document}
  
\newpage

\begin{center}
 {\Large\bf Appendix}
\end{center}
\thispagestyle{empty}
\section*{Proof of Lemma \ref{lmx}}
 In {\bf Configuration~1} (resp. {\bf Configuration~2}), the result follows from the 
 equality of the distance of the intersecting points $i_1$ and $i_2$ from
 $f$ and $h$ (resp. $e$ and $g$), respectively.

\section*{Proof of Theorem \ref{convex_polygon}}

The correctness of the algorithm follows from the facts that (i) we have only two 
configurations of the optimum solution (see Observation \ref{ob}), (ii) in 
Configuration 1, for every point $\theta$ in the left-partition (resp. right-partition) 
$d_\infty(\theta,h)$ $<$ (resp.~$>$) $d_\infty(\theta,f)$ (similar things in Configuration 2), and (iii) 
we are covering portions of the members in $\cal L$ in the left (resp. right) partition by ${\cal S}_1$
(resp. ${\cal S}_2$).

As mentioned in the algorithm, we scanned the input array only twice in 
sequential manner. For each input element, we computed $\theta$ for both 
{\bf Configuration-1} and {\bf Configuration-2} in $O(1)$ time, and updated 
$\sigma_1$ and $\sigma_2$ if needed. The extra space required is to store 
all the variables $e$, $f$, $g$, $h$, $\lambda_1$, $\lambda_2$, $\sigma_1$, 
$\sigma_2$, $\delta$ is $O(1)$. 

\section*{Proof of Lemma \ref{lower_bound_lm}}

Let $r^* < r'$, $D_1$ and $D_2$ (of radius $r'$) be the in-circles of ${\cal S}_1$ and 
${\cal S}_2$. Let $D_3$ and $D_4$ (of radius $r^*$) be the optimum solution for the 
two center problem, and $S_3$ and $S_4$ be the minimum size 
squares covering/ hitting/ restricted-covering
$D_3$ 
and $D_4$ respectively. Thus, $S_3$ and $S_4$ is also a solution for covering $\cal L$ 
by two axis-parallel congruent squares. Since $r^*< r'$, the size of $S_3$ ($S_4$) is less than the 
size of $S_1$ ($S_2$). This leads to the contradiction that $(S_1,S_2)$ is the optimum 
solution for the LCOVER/ LHIT/ restricted LCOVER problem.  

\end{document}